\documentclass[twocolumn,showpacs,preprintnumbers,amsmath,amssymb,aps,prl,superscriptaddress]{revtex4-1}
\usepackage{color}
\usepackage{graphicx}
\usepackage{textcomp} 
\usepackage{subeqnarray}
\usepackage[caption=false]{subfig} 
\usepackage{ulem}
\usepackage{color}

\usepackage{amsmath}

\begin{document}

\clearpage

\title{Elastic interactions between topological defects in chiral nematic shells}
\author{Alexandre  Darmon}
\affiliation{EC2M, UMR CNRS 7083 Gulliver, ESPCI, PSL Research University, 10 Rue Vauquelin, 75005 Paris, France}
\author{Olivier Dauchot}
\affiliation{EC2M, UMR CNRS 7083 Gulliver, ESPCI, PSL Research University, 10 Rue Vauquelin, 75005 Paris, France}
\author{Teresa Lopez-Leon}
\affiliation{EC2M, UMR CNRS 7083 Gulliver, ESPCI, PSL Research University, 10 Rue Vauquelin, 75005 Paris, France}
\author{Michael Benzaquen}
\affiliation{EC2M, UMR CNRS 7083 Gulliver, ESPCI, PSL Research University, 10 Rue Vauquelin, 75005 Paris, France}
\affiliation{Current address: Capital Fund Management, 23 Rue de l'Universit\'e, 75007 Paris, France}

\date{\today}
\begin{abstract}
We present a novel, self-consistent and robust theoretical model to investigate elastic interactions between topological defects in liquid crystal shells. Accounting for the non-concentric nature of the shell in a simple manner, we are able to successfully and accurately explain and predict the positions of the defects, most relevant in the context of colloidal self-assembly. We calibrate and test our model on existing experimental data, and extend it to all newly observed defects configurations in chiral nematic shells. We perform new experiments to check further and confirm the validity of the present model. Moreover, we are able to obtain quantitative estimates of the energies of $+1$ or $+3/2$ disclination lines in cholesterics, whose intricate nature was only reported recently.

\end{abstract}
\maketitle

Topological defects are a common feature of many forms of condensed matter \cite{Mermin1979,Nelson2002b}. They are notably encountered in solids, for which they provide very specific electrical and mechanical properties \cite{Mura2013}. Topological defects are also crucial in other fields such as magnetism \cite{Yu2010} or cosmology \cite{Kibble2000}. 
Although the underlying physics is in each case different, the mathematical framework is universal: the defects are defined as singularities in the order parameter field. One of the most common occurrences of topological defects in condensed matter is in liquid crystals \cite{Janich1987,Kleman1989,deGennes1993}, where they have been widely studied since Lehmann's first description of liquid crystalline mesophases \cite{Lehmann1904}.  

One of the simplest ways to stabilize defects in liquid crystals is to induce topological constraints \cite{Lavrentovich1998}. When a two-dimensional nematic phase is coated onto the surface of a sphere, frustrations in the orientational order necessarily stem from curvature and result in the presence of topological defects. In this context, an original idea, proposed by D. R. Nelson \cite{Nelson2002}, was to use spherical nematic particles as mesoscopic atoms. 
The defects could, once functionalized, act as sticky patches able to induce directional bonds between particles.  These anisotropic building blocks are then expected to reproduce crystalline structures at the mesoscale \textit{via} self-assembly. A good control over the valency, \textit{i.e.} the number of defects, and the bond directionality, \textit{i.e.} the position of the defects, is thus crucial in this context. Since Nelson's seminal paper, many experimental studies focused on nematic shells have demonstrated the applicability of such concepts \cite{Fernandez2007,Lopez-Leon2011,Lopez-Leon2012b,Gharbi2013}. Besides, unexpected symmetries/valencies have been recently reported in cholesteric shells \cite{Darmon2016,Darmon2016b}, in which there is a spontaneous helical arrangement of the director field. Remarkably, it is possible to achieve a good control over the equilibrium defect positions by tuning the {shell thickness heterogeneity} \cite{Lopez-Leon2011,Darmon2016,Zhou2016}. This feature could then be further exploited to produce shells with a variable bonding directionality.  Although numerical studies have been able to capture this idea at the qualitative level \cite{Sec2012b,Koning2013,Wand2015}, no theoretical model is yet able to predict the equilibrium defect positions quantitatively, despite the potential interest for applications. 

\begin{figure}[b!]
\begin{center}
\includegraphics[width= 0.9 \columnwidth]{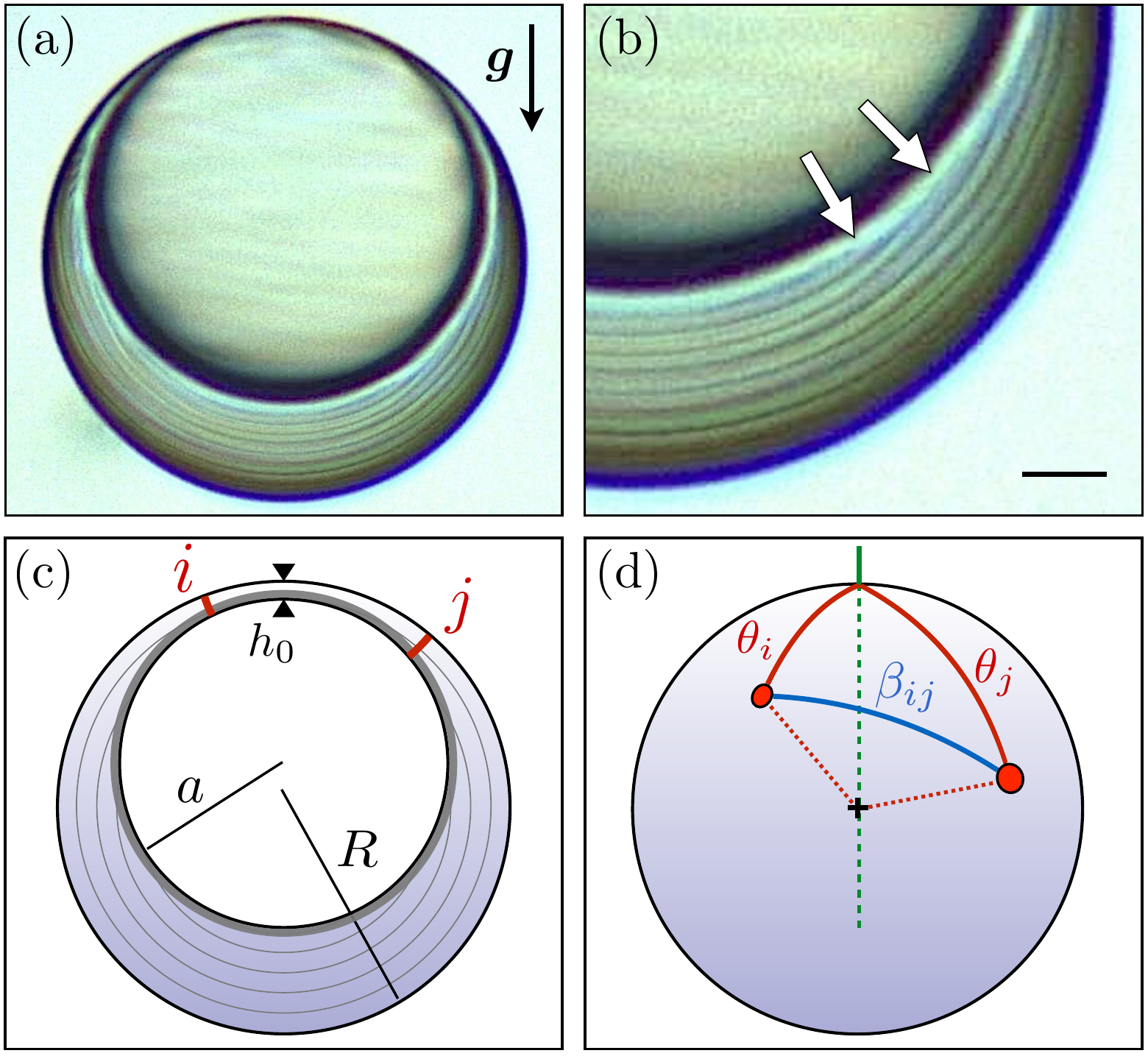}
\end{center}
\caption{(a)-(b) Side view optical microscopy pictures of an eccentric cholesteric liquid crystal shell. Scale bar: $10\,\mu$m. (c) Side view schematics of an eccentric shell. {The concentric cholesteric layers are represented by thin solid grey lines. The thick grey line on the inner sphere signifies the region where the director field becomes slightly distorted to satisfy the tangential anchoring.} (d)~Schematics of defects on the outer sphere.} \label{Side}
\label{schematics}
\end{figure}

In this paper, we present a novel robust and self-consistent approach to compute and predict defect positions in eccentric cholesteric shells (see Fig.~\ref{schematics}).  Minimizing the {free} energy, which we write as surface energy terms multiplied by carefully chosen shell thicknesses, we derive the angular positions of the defects as a function of the shell geometry for all possible defect configurations.  
We first confront our model to available experimental data on the tetravalent configuration, which notably allows us to set the value of the adjustable parameter of our model, namely the minimum shell thickness. We then successively address all other defect configurations together with new experimental data. After performing self-consistency checks,  we use our model to estimate the energies of recently reported non trivial defect structures. \cite{Sec2012,Darmon2016,Darmon2016b}.

On a sphere, a two-dimensional in-plane director field must fulfill the topological requirements of the Poincar\'e-Hopf theorem \cite{Poincare1885,Hopf1926,Kamien2002}. The latter can be written as $\sum_i m_i = 2$, where $m_i$ is the charge or winding number, quantifying the amount of rotation of the director field around defect $i$. Hence, an overall charge of $+2$ needs to be distributed over one or several defects, whose winding numbers are integers or half-integers, consistent with the two-fold symmetry of the nematic phase. Theoretical calculations have shown that the ground state of such a system is tetravalent, composed of four $+1/2$ defects located at the vertices of a regular tetrahedron \cite{Lubensky1992}. Interestingly, this configuration has been experimentally found in nematic and cholesteric shells \cite{Fernandez2007,Darmon2016b}. Besides, four additional configurations with the following defect charges have been recently reported in chiral nematic shells \cite{Darmon2016b}: $(i)$ two $+1$ defects, $(ii)$ one +1 and two +1/2 defects, $(iii)$ one $+3/2$ and one $+1/2$ defects, and $(iv)$ one $+2$ defect. In the following, we investigate each of the above configurations in the stated order for reasons that shall become clear further down this paper.

Thickness heterogeneities in the shell are due to a density mismatch between the inner phase and the liquid crystal phase (see Fig.~\ref{schematics}(a)). 
As suggested in references \cite{Lopez-Leon2011,Koning2013}, we here quantitatively argue that the equilibrium positions of the defects result from a balance of two forces: $(i)$ an elastic repulsion that drives defects away from each other, and $(ii)$ an attractive thickness gradient arising from the non-concentricity of the shells (see Fig.~\ref{Side}). As a result, topological defects tend to regroup in the thinnest part of the shell. Figure~\ref{schematics}(a-b) displays a side imaging of a typical cholesteric shell. The uncompressed cholesteric layers arrange as concentric spheres, starting from the outer surface of the shell, and the observed helical periodicity in the shell matches the actual pitch of the chiral solution, meaning that there is no frustration of the spontaneous cholesteric twist. Due to the presence of the inner droplet and the eccentric nature of the shell, each layer ends at a different position on the inner surface, as indicated by the arrows on Fig.~\ref{schematics}(b).   {Note that this arrangement is, \textit{a priori}, not compatible with a planar degenerate anchoring on the inner surface. Since we use PVA which provides strong tangential anchoring, there actually exists a small region around the inner sphere where the director field becomes slightly distorted to overcome this issue (see Fig.~\ref{schematics}(b)-(c)). However, this surface contribution is \textit{a priori} small compared to the other elastic costs in the system.} 

{In cholesteric shells, the arrangement of the director field can thus be described as concentric layers with a helical twist matching the actual pitch of the cholesteric solution. Remarkably, if such a director field is introduced in the Frank-Oseen free energy density, the twist term vanishes \cite{Bezic1992}. Theoretically, it was even shown that for cholesteric droplets, this director field minimizes the free energy of the system \cite{Bezic1992}. It is precisely the spontaneous cholesteric twist that makes the twist contribution null in these spherical systems.}
At the first level of approximation, we can thus ignore the details of the molecular ordering. In particular, since the spontaneous helical  pitch is everywhere satisfied in this geometry, we consider the global cholesteric arrangement as a superposition of two-dimensional nematic layers. {Interestingly, the above-mentioned onion-like arrangement ensured by the spontaneous twist is not necessarily present in nematic shells where the director field can have a non-negligible radial component \cite{Vitelli2006,Koning2013}. For this reason, cholesteric shells are more adapted to our approach than their nematic counterparts.}

In the one elastic constant approximation, the surface {free} energy $E$ of a two-dimensional in-plane director field with topological defects interacting on a sphere can be written as \cite{Nelson2002}: 
\begin{eqnarray}
E &=& \pi K \bigg(  \sum_i E_{i}^0 +\sum_{i<j} U_{ij} \bigg) \ , \label{energy2DNelson}
\end{eqnarray}
where $K$ is the elastic constant, $E_{i}^0$ the dimensionless energy of defect $i$, and $U_{ij}$  the dimensionless interaction energy between defects $i$ and $j$, with:
\begin{subequations}
\begin{align}
E_{i}^0 &=  m_i^2 \log \left(R/r_{\text{c},i} \right) \label{Eci} \\
U_{ij} &= - m_im_j\log\left(1-\cos\beta_{ij}\right)  \label{Eintij} \ ,
\end{align}
\end{subequations}
where $R$ is the sphere radius, $r_{\text{c},i}$ the defect core radius, and $\beta_{ij}$ the central angle between defects $i$ and $j$ (see Fig.~\ref{schematics}(d)). For small angles, Eq.~\eqref{Eintij} reduces to:
\begin{eqnarray}
U_{ij} &=& -2 m_im_j\log\left(\beta_{ij}/\sqrt{2}\right)  \ .\label{Eintij2}
\end{eqnarray}
In order to compute the total {free} energy of the system and account for the eccentric nature of the shell, we proceed as follows. Rather than performing a highly nontrivial integration of the energy over the non-concentric system, we aim at capturing the essence of the interaction in a simple effective way. To do so, we multiply the different terms in the two-dimensional {free} energy of Eq.~\eqref{energy2DNelson} by local shell thicknesses taken according to the physical grounds provided below. In the following, these thicknesses are expressed in units of the outer radius of the shell $R$. For the self part of the energy $E_{i}^0$, we shall simply use the local thickness of the shell at defect~$i$, denoted $h_i$. 
For the interaction energy, we use the minimal thickness along the geodesic path between defects $i$ and $j$ on the outer sphere, denoted $h^{\text{min}}_{ij}$. The reason is as follows. As mentioned above, the inner water droplet does not compress the cholesteric layers which, on the contrary, are interrupted at its boundary (see Fig.~\ref{schematics}(a-b)). We thus assume {that the defect interaction is mostly mediated by the elastic energy of the layers that are not disrupted by the inner droplet}, the extension of those layers being proportional to $h^{\text{min}}_{ij}$.
The total dimensionless {free} energy $F=E/(\pi K R)$ of the eccentric liquid crystal shell can then be written as:
\begin{equation}
F =  \sum_i E^0_{i} \, h_i + \sum_{i<j} U_{ij} \, h^{\text{min}}_{ij}  \ .
\label{Fgeneral}
\end{equation}

Multiplying by local thicknesses is at the core of the present model, and corresponds to the simplest approach where the attractive thickness gradient is taken into account. 
The approach is thus expected to be most accurate when the defects are close to each other. 

Due to the azimuthal symmetry of the eccentric shell, the local thickness of the shell denoted $h$ is a function of the polar angle $\theta$ only (see Fig.~\ref{schematics}(d)). As a result, the thicknesses involved in Eq.~\eqref{Fgeneral} are such that $ h_i =  h(\theta_i)$, where $\theta_i$ denotes the polar angle of defect $i$, and $ h^{\text{min}}_{ij} =  h(\theta^{\text{min}}_{ij})$,  where $\theta^{\text{min}}_{ij}$ is the polar angle where $h$ is minimal  along the geodesic path between $i$ and $j$.  When varying the geometry of the shell, $h(\theta)$ is also an implicit function of two additional dimensionless parameters: $(i)$ the renormalized minimal thickness $h_0$, and $(ii)$ $u\equiv(R-a)/R$, where $a$ denotes the inner radius of the shell (see Fig.~\ref{schematics}(c)). 
In our experiments, we observe that $h_0$ is constant and independent of the shell nature and geometry. On physical grounds, this can be explained by the fact that it is the disjoining pressure between the inner and outer interfaces which sets the value of  $h_0$ \cite{Darmon2016} (see \cite{Zhou2016} for a recent numerical study on the effects of varying $h_0$).
Hence, we are only left with the parameter $u$, which actually measures the thickness gradient within the shell, and shall thus rigorously write $h(\theta;u)$ for the local thickness. 
Finally, for each of the defect configurations, we minimize the total {free} energy $F$ (see Eq.~\eqref{Fgeneral}) with respect to the  angular positions, and obtain the equilibrium angles as function of $u$ only. In the small angle approximation, one can easily show using purely geometrical considerations that the local thickness $ h(\theta;u)$ reads:
\begin{equation}
 h(\theta;u) =  h_0 + g(u)\,\frac{\theta^2}{2} +\text{o}\left(\theta^2\right) \ , \label{hpetittheta}
 \end{equation}
 where $g$ is a dimensionless function of $u$ only reading:
 \begin{equation}
g(u)=\frac{(1-{h}_0)(u-{h}_0)}{1-u} \ . \label{g(u)}
 \end{equation}
Let us start with the tetravalent $4[+1/2]$ configuration. To confront our model, we use experimental data of nematic shells from reference \cite{Lopez-Leon2011}. The reasons for such a choice are two-fold. First, all defects in this configuration are singular lines such that the arrangement of the director field remains essentially two-dimensional. As mentioned above, this feature is crucial in our approach. Second, the exact structure of those lines is well known which, as we shall see below, is not always the case in cholesterics. This configuration is thus the best candidate to check the validity of the present model.
 \begin{figure}[t!]
\begin{center}
\includegraphics[width= 1. \columnwidth]{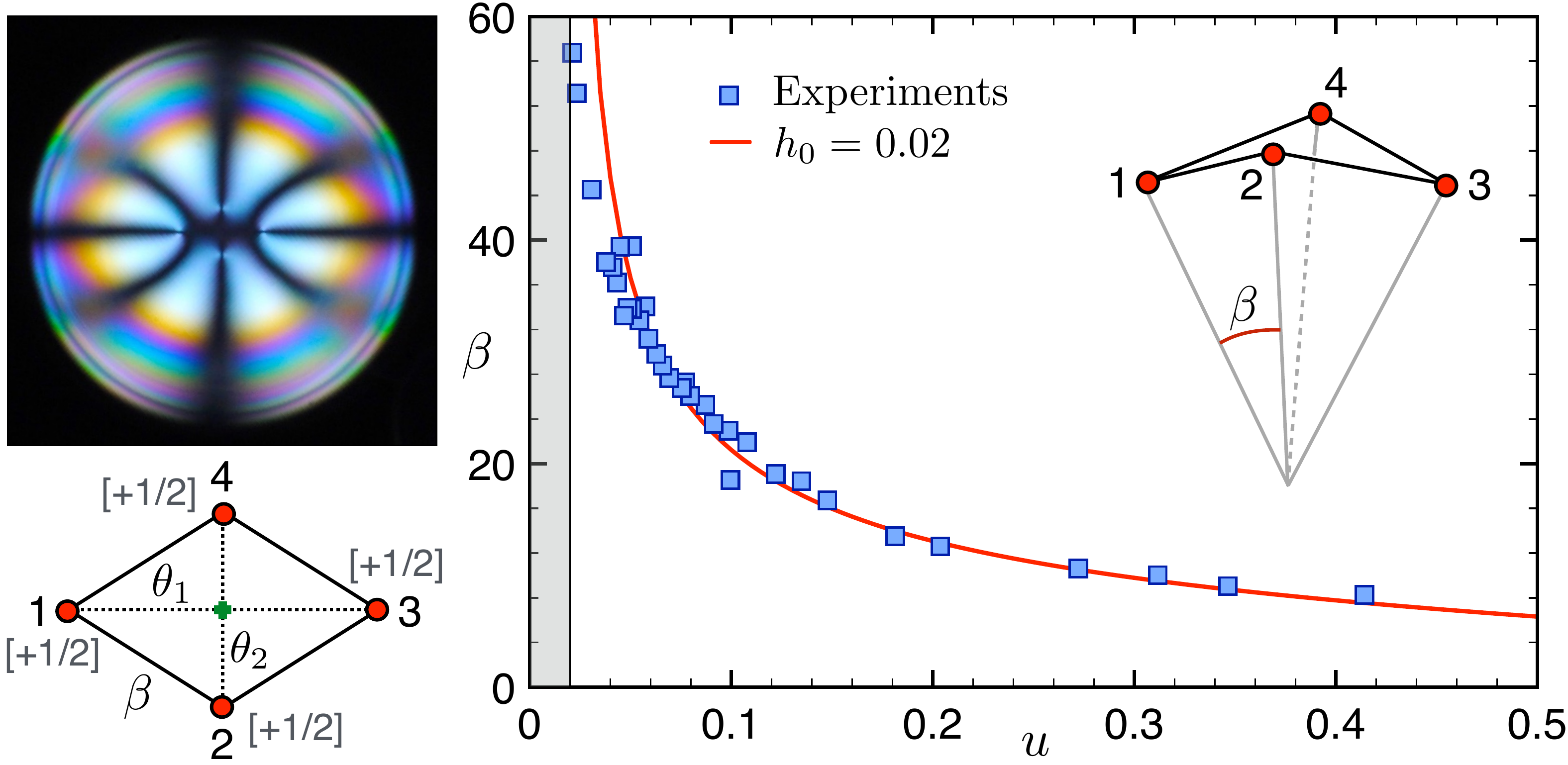}
\end{center}
\caption{Configuration with four +1/2 defects. Top left: Top view image between crossed polarizers of a $4[+1/2]$ nematic shell. Bottom left: Schematics of defects arrangement. Right: Angular distance $\beta$ between nearby defects  as function of $u$. {The blue squares correspond to experimental data from \cite{Lopez-Leon2011} and the solid red line is the result of the minimization of the free energy in Eq.~\eqref{EnergyTetra} with $h_0=0.02$}.} \label{Tetra}
\label{tetra}
\end{figure}
It is notably characterized by four outer defects located at the vertices of a folded rhombus (see Fig.~\ref{Tetra}). The experimental central angle $\beta$ between two nearby defects, identical for each pair of defects {and taken from reference \cite{Lopez-Leon2011}}, is plotted as a function of $u$. Noting that $\theta_1=\theta_3$, $\theta_2=\theta_4$, and that $ h(\theta^{\text{min}}_{13};u)= h(\theta^{\text{min}}_{24};u)= h_0$, the {free} energy of the 4[+1/2] configuration reads:
\begin{equation}
\begin{split}
 F_{4[+1/2]} (\theta_1,\theta_2;u)=& \  4 U_{12} \,  h(\theta^{\text{min}}_{12};u) + ( U_{13}+ U_{24}) \,  h_0\\
&+ 2 E^0_{1} \, \left[ h(\theta_1;u)+ h(\theta_2;u)\right] \ , \label{EnergyTetra}
\end{split}
\end{equation}
where the interaction energies read:
\begin{subeqnarray}
 U_{12}(\theta_1,\theta_2)&=& - \frac 14 \log\left(\frac{\theta_1^2+\theta_2^2}{2} \right)  \\
 U_{13}(\theta_1)&=& -  \frac 12 \log\left(\theta_1\sqrt{2} \right)\\
 U_{24}(\theta_2)&=& - \frac 12 \log\left(\theta_2\sqrt{2} \right) \ ,
\end{subeqnarray}
and where the angle $\theta^{\text{min}}_{12}$ in Eq.~\eqref{EnergyTetra} is given by:
\begin{eqnarray}
\theta^{\text{min}}_{12} &=& \frac{\theta_1 \theta_2}{\sqrt{\theta_1^2+\theta_2^2}} \ .
\end{eqnarray}
The two parameters $\theta_1$ and $\theta_2$ fully characterize the positions of the defects.. We set $r_{\text{c},1/2} \sim10$\,nm $\sim 10^{-4} R$ for $R = 100 \,\mu$m, consistently with reported values \cite{deGennes1993,Oswald2005}. Minimizing the {free} energy with respect to $\theta_1$ and $\theta_2$, and noting that $\beta =\sqrt{\theta_1^2 + \theta_2^2}$, we obtain the equilibrium curve $\beta(u)$ (see Fig.~\ref{tetra}). Fitting the experimental data to $\beta(u)$ with respect to $h_0$ yields excellent agreement for $h_0=0.02$. The latter value of $h_0$, equal to $1\,\mu m$ when $R=50\,\mu$m, is consistent with the current and previous experimental studies \cite{Lopez-Leon2011,Darmon2016}. This first result can be seen as a calibration of the model and we shall use the above value of $ h_0$ as a reference throughout the following.\\

 \begin{figure}[t!]
\begin{center}
\includegraphics[width= 1 \columnwidth]{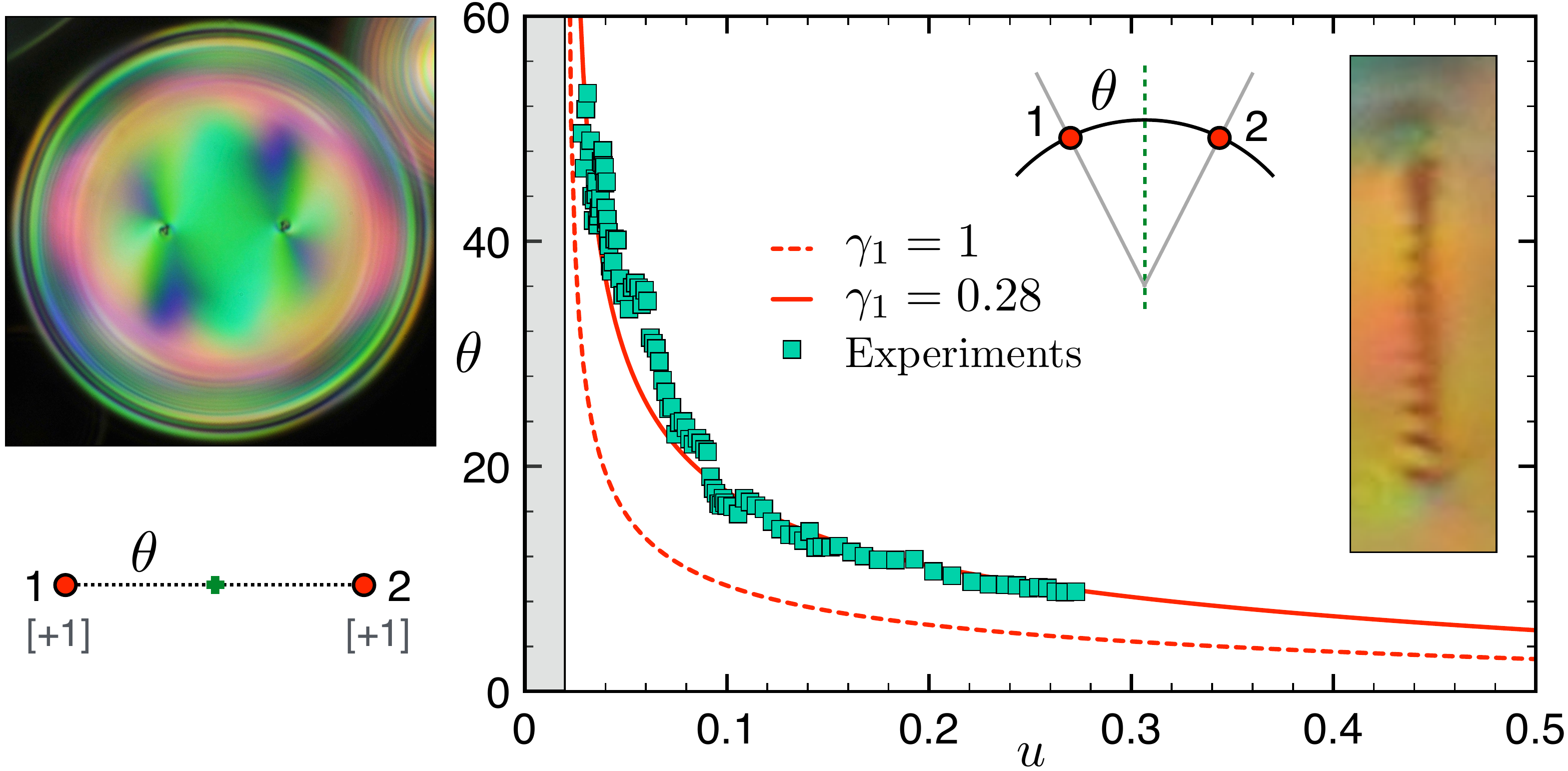}
\end{center}
\caption{Configuration with two +1 defects. Top left: Top view image between crossed polarizers of a $2[+1]$ cholesteric shell. Bottom left: Schematics of defects arrangement. Right: Angular position $\theta$ as function of $u$. Experimental data from \cite{Darmon2016} for which a rolling average was performed. The inset displays a picture of the intricate structure of the $+1$ disclination in a cholesteric shell.} \label{BivDSS}
\label{trajectories}
\end{figure}
 
We now look into the configuration consisting of two +1 disclination lines. In Fig.~\ref{BivDSS}, we report data obtained from a previous study \cite{Darmon2016}, measured for shells with different cholesteric pitches $p$ = 9.3, 6, and 3.6\,$\mu$m (see green squares in Fig.~\ref{BivDSS}). Noting that $\theta_1=\theta_2\equiv\theta$, and that $ h(\theta^{\text{min}}_{12};u)= h_0$, the {free} energy of the $2[+1]$ configuration reads:
\begin{equation}
 F_{2[+1]}(\theta;u) = \,  U_{12} \,  h_0 + 2 E^0_{1} \,  h(\theta;u) \ ,
\label{EnergyBivDSS}
\end{equation}
where the interaction energy reads:
\begin{eqnarray}
 U_{12}(\theta)&=&  -2 \log\left(\theta \sqrt{2} \right) \ .
\end{eqnarray}
For this configuration as well as for the following, the  core radii are set according to the reference value $r_{\text{c},1/2}$ through $r_{\text{c},i} = (m_i/m_j)  r_{\text{c},j}$ \cite{Oswald2005}. Minimization with respect to $\theta$ yields the red dashed line in Fig.~\ref{BivDSS}, which does not quantitatively capture the experimental behavior. The reason is that $+1$ lines in cholesteric shells are actually not simple singular lines.
Our recent experiments and numerical simulations \cite{Darmon2016} have shown that this structure is actually composed of a stack of disclination rings (see inset of Fig.~\ref{trajectories}), with a director field escaping between each ring. {The energy of such a defect is thus expected to be different from that of a purely singular line. The proper self-energy shall then be written as $ \gamma_1 E^0_{1}$, where $ \gamma_1$ is a scalar which is \textit{a priori} unknown. It is worth emphasizing here that only the self-energy of the defect is altered. Indeed, although the structure of the defect is intricate, it can still be seen as a +1 disclination when looking sufficiently far off. Hence, since the interaction energy is far field, it should not be affected by the details of the defect structure.}  Best fit is obtained for $\gamma_1=0.28$, displayed as a solid red line in 
Fig.~\ref{BivDSS}. {Note that we obtain $\gamma_1<1$ which is fully consistent with the escaped nature of the disclination.} It is worth mentioning that this approach, which we shall self-consistently validate below, actually represents a novel method to experimentally estimate the energy of this intricate structure. \\

\begin{figure}[t!]
 \centering
\includegraphics[width= 1 \columnwidth]{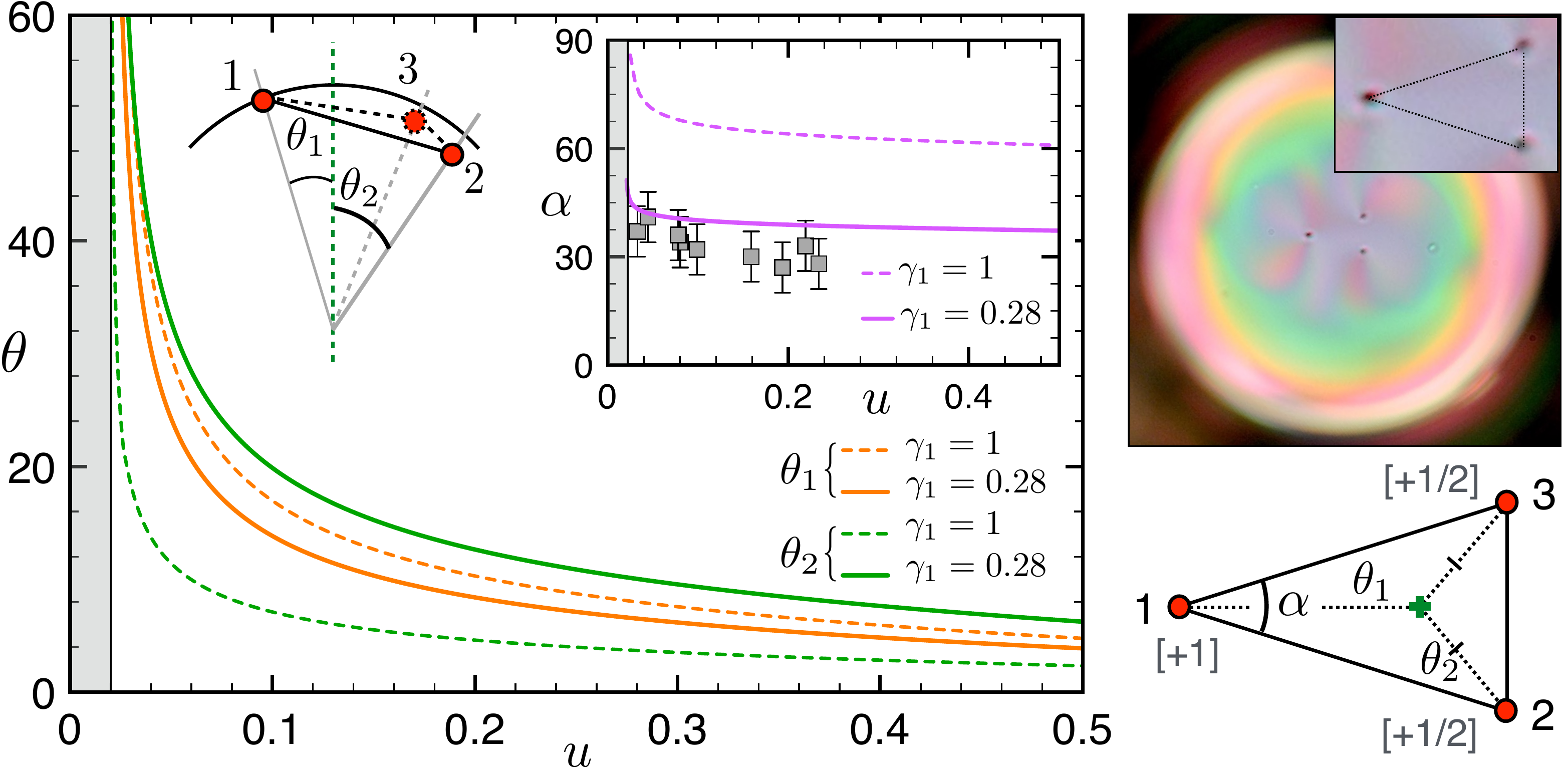}
\caption{Configuration with two +1/2  and one +1 defects. Left: Angular positions $\theta_i$  as function of $u$. The inset displays the vertex angle $\alpha $ of the isosceles triangle as function of $u$ (experiments in grey squares). Top right: Top view image between crossed polarizers of a $[+1]+ 2[+1/2]$ cholesteric shell with a zoom on the defects (inset). Bottom right: Schematics of defects arrangement.}
 \label{triangle}
\end{figure}

Having addressed configurations made of $+1/2$ and $+1$ defects only, the next logical step consists in studying a configuration containing both, namely the triangular configuration $[+1]+ 2[+1/2]$. To fully parametrize the positions of the defects on the outer sphere, three independent parameters, namely $\theta_1$, $\theta_2$ ($=\theta_3$), and $\alpha$, which is the vertex angle of the isosceles triangle, must be considered (see Fig.~\ref{triangle}). 
The dimensionless {free} energy can be written as:
\begin{equation}
\begin{split}
 F_{\text{triangle}}(\theta_1,\theta_2,\alpha;u) =& \  2 U_{12} \,  h(\theta_{12}^{\min};u) +  U_{23} \,  h(\theta_{23}^{\min};u)\\
&+\gamma_1   E^0_{1} \,  h(\theta_1;u)+ 2 E^0_{2} \,  h(\theta_2;u)\ .
\end{split}
\label{EnergyTriangle}
\end{equation}
where the interaction energies in Eq.~\eqref{EnergyTriangle} read:
\begin{eqnarray}
 U_{12}(\theta_1,\theta_2,\alpha)&=& -  \log\left(\frac{\beta_{12}}{\sqrt{2}} \right)  \\
 U_{13}(\theta_1,\theta_2,\alpha)&=& -  \frac 12 \log\left(\frac{\beta_{23}}{\sqrt{2}} \right) \ ,
\end{eqnarray}
with central angles reading:
\begin{align}
&\beta_{12} = \theta_1 \cos\left(\frac{\alpha}{2}\right) +  \sqrt{\theta_2^2-\theta_1^2 \sin^2\left(\frac{\alpha}{2}\right)} \\
&\beta_{23} =2 \sin\left(\frac{\alpha}{2}\right)\left(\theta_1 \cos\left(\frac{\alpha}{2}\right) + \sqrt{\theta_2^2-\theta_1^2 \sin^2\left(\frac{\alpha}{2}\right)}  \right) \ ,
\end{align}
and where the angles $\theta^{\text{min}}_{12}$ and $\theta^{\text{min}}_{32}$ in Eq.~\eqref{EnergyTriangle} read:
\begin{align}
&\theta_\text {12}^{\min} = \cos\left(\frac{\alpha}{2}\right)\left(\theta_1 \cos\left(\frac{\alpha}{2}\right) + \sqrt{\theta_2^2-\theta_1^2 \sin^2\left(\frac{\alpha}{2}\right)}  \right) - \theta_1 \\
&\theta_\text {23}^{\min} = \frac 12 \tan\left(\frac{\alpha}{2}\right)\left(\theta_1 \cos\left(\frac{\alpha}{2}\right) + \sqrt{\theta_2^2-\theta_1^2 \sin^2\left(\frac{\alpha}{2}\right)}  \right) \ . 
\end{align}
One shall note that the self-energy of the $+1$ defect has naturally been set to $\gamma_1   E^0_{1}$, consistent with the  2[+1] case, and where $E^0_{2}$ denotes the self energy of the $+1/2$ defects. The equilibrium solutions $\theta_1$ and $\theta_2$ are displayed in Fig.~\ref{triangle}, both for  $\gamma_1=1$ (dashed lines) and $\gamma_1=0.28$ (solid lines). 
Most importantly, when plotting the vertex angle $\alpha$ as a function of $u$ (see inset of Fig.~\ref{triangle}),  we see that this angle is actually much larger for $\gamma_1=1$ than for $\gamma_1=0.28$. To confront this theoretical prediction on the vertex angle to experimental data, we have generated cholesteric shells using microfluidics \cite{Utada2005,Fernandez2007}, and measured $\alpha$ for different cholesteric shells (see grey squares in the inset of Fig.~\ref{triangle}). One can see that the model with $\gamma_1=0.28$ is clearly better than with $\gamma_1=1$. The results on the triangle configuration constitute a self-consistency check that validates the model and confirms the estimate of the energy found for the intricate +1 disclination in cholesteric shells.

\begin{figure}[t!]
 \centering
 \includegraphics[width= 1 \columnwidth]{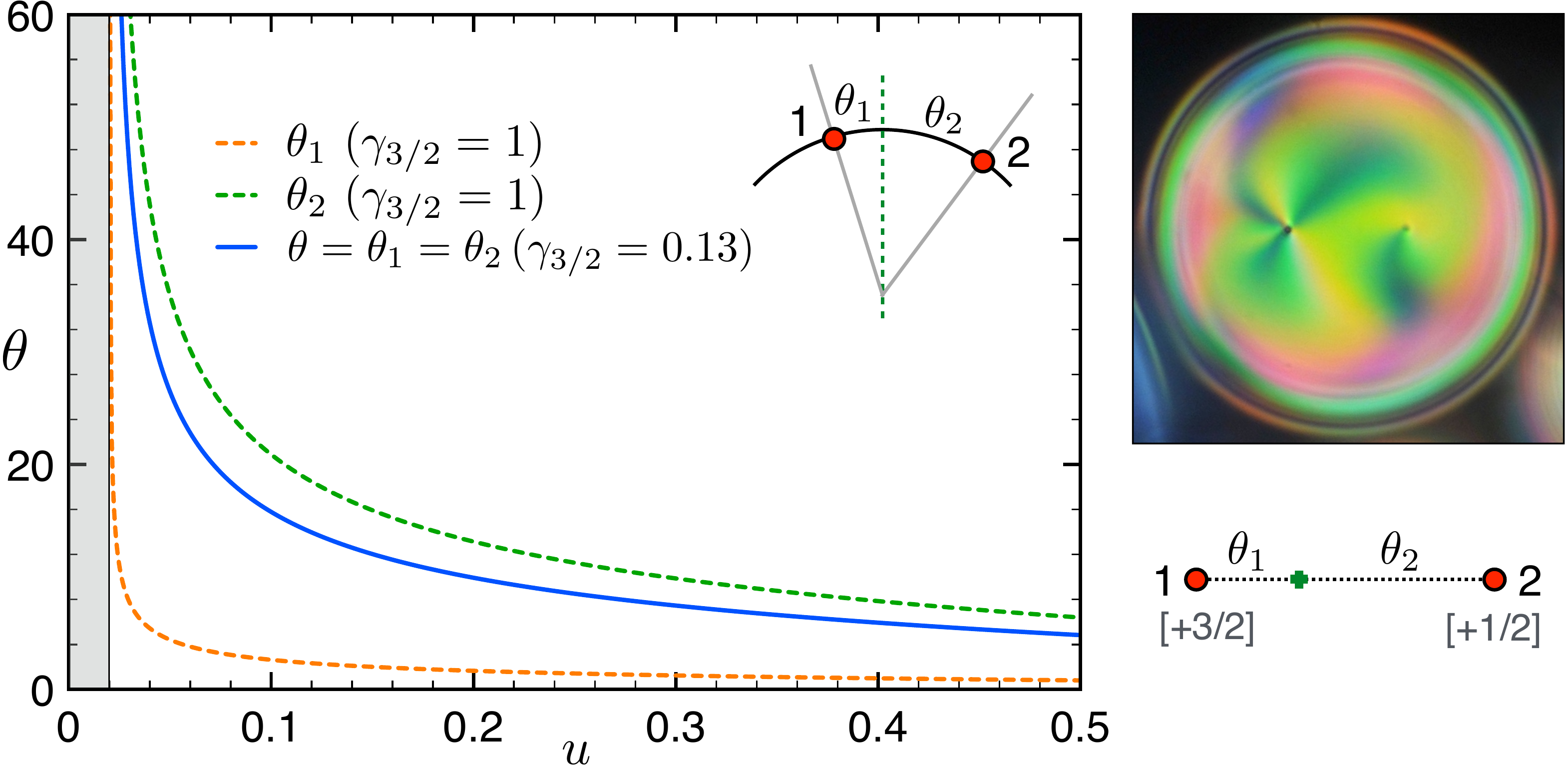}
 \caption{Configuration with one +3/2 and one +1/2 defects. Left: Angular positions $\theta_i$ as function of $u$. Top right: Top view image between crossed polarizers of a $[+3/2] + [+1/2]$ cholesteric shell. Bottom right: Schematics of defects arrangement.}
 \label{threehalf}
\end{figure}

Let us finally investigate the last and newly reported configuration, composed of two defects of non equal charges +3/2 and +1/2 \cite{Darmon2016b}. In the following, the indices 1 and 2 respectively correspond to the +3/2 and +1/2 defects (see Fig.~\ref{threehalf}). In this configuration, the dimensionless {free} energy reads: 
\begin{equation}
\begin{split}
 F_{[+3/2]+[+1/2]}(\theta_1,\theta_2;u) = & \  U_{12} \,  h_0 +  E^0_{1} h(\theta_1;u) \\
 &+   E^0_{2} \,  h(\theta_2;u) \ ,
\label{EnergyThreehalf}
\end{split}
\end{equation}
where the interaction energy reads:
\begin{eqnarray}
 U_{12}(\theta_1,\theta_2)&=&  -\frac 32 \log\left(\frac{\theta_1+\theta_2}{ \sqrt{2}} \right) \ .
\end{eqnarray}
The equilibrium angles $\theta_1$ and $\theta_2$ are plotted as a function of $u$ on Fig.~\ref{EnergyThreehalf} (dashed lines). As expected, since $E^0_{1}$ is much larger than $E^0_{2}$,  $\theta_1$ is much smaller than $\theta_2$ for all $u$. 
However, our experiments reveal that $\theta_{1,\text{exp}} \simeq \theta_{2,\text{exp}}$ for all $u$, meaning that the true energy of the $+3/2$ should be very much comparable to that of the $+1/2$. Hence, the energy of the +3/2 disclination must be corrected by a factor denoted $\gamma_{3/2}$. This necessary correction is, here as well, consistent with the intricate structure of the non regular +3/2 defect line, actually made of a non-singular disclination wound around another singular line \cite{Darmon2016b}. To match the experimental observations, the energy of the real $+3/2$ line must naturally be equal to that of the $+1/2$ line; one finds: $\gamma_{3/2} =  E^0_{2} /  E^0_{1} \simeq 0.13$ (see solid blue line in Fig.~\ref{threehalf}). 

Finally, the single $+2$ defect equilibrium configuration is trivial, as there is no interaction term $U_{ij}$; it is naturally always located at $\theta=0$.

We have shown that our self-consistent model is able to successfully explain and predict defect positions in chiral nematic shells. In the context of colloidal self-assembly, fine tuning of defect positions is of crucial importance as it controls the bond directionality of these super-atom candidates \cite{Nelson2002,Lopez-Leon2011}. Moreover and equally important, our model allows to estimate the energies of the recently reported highly nontrivial structures displayed by cholesterics in spherical geometries.  More generally, the framework developed in this letter opens the way to a novel method to measure unknown energies of defect cores.

We thank A. Fernandez-Nieves for fruitful discussions. We acknowledge  support from Institut Pierre-Gilles de Gennes (laboratoire d'excellence, Investissements d'avenir program ANR-10-IDEX 0001-02 PSL and ANR-10-EQPX-31), as well as the ANR with grant number 13-JS08-0006-01.

\bibliographystyle{h-physrev}
\bibliography{biblio}

\begin{thebibliography}{10}

\bibitem{Mermin1979}
N.~D. Mermin,
\newblock Rev. Mod. Phys. {\bf 51}, 591 (1979).

\bibitem{Nelson2002b}
D.~R. Nelson,
\newblock {\em Defects and Geometry in Condensed Matter Physics} (Cambridge
  University Press, 2002).

\bibitem{Mura2013}
T.~Mura,
\newblock {\em Micromechanics of defects in solids} (Springer Science and
  Business Media, 2013).

\bibitem{Yu2010}
X.~Z. Yu {\em et~al.},
\newblock Nature {\bf 465}, 901 (2010).

\bibitem{Kibble2000}
T.~W.~B. Kibble,
\newblock {\em Classification of Topological Defects and their Relevance to
  Cosmology and Elsewhere. In Topological Defects and the Non-Equilbrium
  Dynamics of Symmetry Breaking Phase Transitions (eds Bunkov, Y. M. and
  Godfrin H.)} (NATO Science Series, Series C: Mathematical and Physical
  Sciences, 2000).

\bibitem{Janich1987}
K.~J\"anich,
\newblock Acta Applic. Math {\bf 8}, 65 (1987).

\bibitem{Kleman1989}
M.~Kl\'eman,
\newblock Rep. Prog. Phys. {\bf 52}, 555 (1989).

\bibitem{deGennes1993}
P.-G. de~Gennes and J.~Prost,
\newblock {\em The Physics of Liquid Crystals}, 2nd ed. (Oxford University
  Press, 1993).

\bibitem{Lehmann1904}
O.~Lehmann,
\newblock {\em Fl\"{u}ssige Kristalle} (Wilhelm Engelmann, Leipzig, 1904).

\bibitem{Lavrentovich1998}
O.~D. Lavrentovich,
\newblock Liquid Crystals {\bf 24}, 117 (1998).

\bibitem{Nelson2002}
D.~R. Nelson,
\newblock Nano Lett. {\bf 2}, 1125 (2002).

\bibitem{Fernandez2007}
A.~Fernandez-Nieves {\em et~al.},
\newblock Phys. Rev. Lett. {\bf 99}, 157801 (2007).

\bibitem{Lopez-Leon2011}
T.~Lopez-Leon, V.~Koning, K.~B.~S. Devaiah, V.~Vitelli, and
  A.~Fernandez-Nieves,
\newblock Nature Phys. {\bf 7} (2011).

\bibitem{Lopez-Leon2012b}
T.~Lopez-Leon, M.~Bates, and A.~Fernandez-Nieves,
\newblock Phys. Rev. E {\bf 86}, 030702 (2012).

\bibitem{Gharbi2013}
M.~A. Gharbi {\em et~al.},
\newblock Soft Matter {\bf 9}, 6911 (2013).

\bibitem{Darmon2016}
A.~Darmon, M.~Benzaquen, O.~Dauchot, and T.~Lopez-Leon,
\newblock Proc. Natl. Acad. Sci. {\bf 113}, 9469 (2016).

\bibitem{Darmon2016b}
A.~Darmon, M.~Benzaquen, S.~\v{C}opar, O.~Dauchot, and T.~Lopez-Leon,
\newblock arXiv:1608.00090  (2016).

\bibitem{Zhou2016}
Y.~Zhou {\em et~al.},
\newblock arXiv:1607.08263  (2016).

\bibitem{Sec2012b}
D.~Se\u{c} {\em et~al.},
\newblock Phys. Rev. E {\bf 86}, 020705 (2012).

\bibitem{Koning2013}
V.~Koning, T.~Lopez-Leon, A.~Fernandez-Nieves, and V.~Vitelli,
\newblock Soft Matter {\bf 9}, 4993 (2013).

\bibitem{Wand2015}
C.~R. Wand and M.~A. Bates,
\newblock Phys. Rev. E {\bf 91}, 012502 (2015).

\bibitem{Sec2012}
D.~Se\u{c}, T.~Porenta, M.~Ravnik, and S.~\u{Z}umer,
\newblock Soft Matter {\bf 8}, 11982 (2012).

\bibitem{Poincare1885}
H.~Poincar\'e,
\newblock J. Math. Pure Appl. {\bf 1}, 167 (1885).

\bibitem{Hopf1926}
H.~Hopf,
\newblock Math. Ann. {\bf 96}, 427 (1926).

\bibitem{Kamien2002}
R.~D. Kamien,
\newblock Rev. Mod. Phys. {\bf 75}, 953 (2002).

\bibitem{Lubensky1992}
T.~Lubensky and J.~Prost,
\newblock J. Phys II France {\bf 2}, 371 (1992).

\bibitem{Bezic1992}
J.~Bezi\'{c} and S.~\u{Z}umer,
\newblock Liquid Crystals {\bf 11}, 593 (1992).

\bibitem{Vitelli2006}
V.~Vitelli and D.~R. Nelson,
\newblock Phys. Rev. E {\bf 74}, 021711 (2006).

\bibitem{Oswald2005}
P.~Oswald and P.~Pieranski,
\newblock {\em Nematic and cholesteric liquid crystals} (Taylor and Francis,
  2005).

\bibitem{Utada2005}
A.~S. Utada {\em et~al.},
\newblock Science {\bf 308}, 537 (2005).

\end{thebibliography}

\end{document}